\documentclass[aps,prd,twocolumn,showpacs,nofootinbib,longbibliography,10pt]{revtex4-1}%
\usepackage[colorlinks,citecolor=DarkGreen,linkcolor=DarkRed,urlcolor=DarkBlue]{hyperref}
\usepackage{amsmath}
\usepackage{graphicx}
\usepackage{slashed}
\usepackage[svgnames]{xcolor}

\begin{document}

\title{Axial coupling constant in a magnetic background}
\author{Cristi\'an Villavicencio}
\affiliation{Centro de Ciencias Exactas \& Departamento de Ciencias B\'asicas, Facultad de Ciencias, Universidad del Bio-Bio,
Casilla 447, Chill\'an, Chile.}
\email{cvillavicencio@ubiobio.cl}

\begin{abstract}
The nucleon axial coupling constant $g_A$ is calculated in the presence of an external uniform magnetic field using finite energy sum rules.
The correlation function of proton, neutron and axial-vector currents is calculated both the hadronic and the QCD sector.
Once the axial contribution in the form factor is isolated, a double sum rule is considered, i.e., the usual QCD contour integration for the external moments of both the proton and neutron current.
The effects of the external magnetic field come mainly through the in-medium current-nucleon coupling constants $\lambda_N(B)$ and the hadronic thresholds $s_0(B)$ provided by the nucleon-nucleon correlators.
As a result, the axial coupling constant decreases in the presence of the magnetic field.
\end{abstract}


\maketitle
The axial-vector form factor is present in weak processes such as beta decay.
In particular, the neutron decay width is proportional to the factor $1+3g_A^2$, with $g_A$ being the axial coupling constant of the nucleon, also known as axial charge, which is experimentally measured from beta decay \cite{Czarnecki:2018okw,Mund:2012fq}, with $g_A \approx 1.275$.
Medium effects on the axial coupling constant could have interesting implications, in particular high density effects in compact stars, where the neutrino emissivity in the Urca process, the main cooling process of neutron stars, is also proportional to the aforementioned factor \cite{Yakovlev:2000jp}.
It is established that axial coupling exhibits quenching, where $g_A \to 1$ for high baryon density \cite{Carter:2001kw}, although non-quenching behavior \cite{Li:2018ykx} has also been suggested.
Thermal effects on the axial coupling constant have also been studied in the framework of relativistic heavy ion collision (HIC) experiments.
In this case, the axial coupling decreases dramatically near the deconfinement/chiral phase transition \cite{Dominguez:1999ka,Abu-Shady:2012ewe}.

One of the important effects to take into account are the strong magnetic fields produced in peripheral HIC experiments, as well as the strong magnetic field inside the magnetars.
In this work, the axial coupling constant in the presence of a uniform and constant external magnetic field will be obtained using finite energy sum rules (FESR).
The series expansion of the magnetic field can be applied in sum rules \cite{Ioffe:1983ju,Cho:2014exa,Ayala:2015qwa,Gubler:2015qok,Dominguez:2018njv,Dominguez:2020sdf}, reaching more realistic values for the magnetic field compared to other methods using Landau level summation.

This paper begins with the proton-axial-neutron currents correlator, looking at how to isolate the axial component in the form factor.
A double FESR is then applied in vacuum. In combination with the results obtained in the nucleon-nucleon current correlator, the axial coupling constant is obtained as a function of the quark condensate and the gluon condensate, which must be properly fixed. 
Magnetic field effects are then introduced using nucleon-current couplings and hadronic thresholds, both dependent on the magnetic field. 
The paper concludes with a discussion of the results and perspectives.

\section{Current correlator}

The starting point is the analysis of the three currents correlator in configuration space,
\begin{equation}
 \Pi_\mu(x,y,z)=-\langle 0|\,{\cal T}\,\eta_p(x)A_\mu(y)\,\bar\eta_n(z)\,|0\rangle, 
\end{equation}
being $\eta_N$ the interpolating current of the nucleons and $A_mu$ the axial-vector current.
In the hadronic sector, the nucleonic and axial currents are defined by 
\begin{align}
 \langle 0|\,\eta_p(x)\,|p',s'\rangle &= \lambda_p\, u_p^{s'}(p')\,e^{-ip'\cdot x},\label{eq.eta_p} \\
 \langle p,s|\,\bar\eta_n(z)\, |0\rangle &= \lambda_n\, \bar u_n^s(p)\, e^{ip\cdot z}\\
 \langle p',s'|A_\mu(y)|p,s\rangle &= \bar u_p^{s'}(p')\,T_\mu(q)\,u_n^s(p) \,e^{iq\cdot y},\label{eq.A}
\end{align}
 with $q=p'-p$, and where $\lambda_p$ and $\lambda_n$ are the current-proton coupling and the current-neutron coupling, respectively.
 The function $T$ is defined as 
\begin{equation}
 T_\mu(q) = 
 G_A(t)\gamma_\mu\gamma_5 +G_P(t)\gamma_5 \frac{q_\mu}{2  m_N}
 +G_T(t)\sigma_{\mu\nu}\gamma_5  \frac{q_\nu}{2 m_N},
 \label{T_mu}
\end{equation}
 with $t=q^2$, and where $ m_N$ is the vacuum nucleon mass.
The axial coupling is defined as 
\begin{equation}
g_A\equiv G_A(0).
\end{equation}

 In the QCD sector, the nucleon interpolating currents and the axial vector current are defined in terms of the quark fields as
\begin{align}
 \eta_p(x) &=\epsilon^{abc}\left[u^a(x)^T\,C\gamma^\mu \,u^b(x)\right]\gamma_\mu\gamma_5 \,d^c(x),\\
 \bar\eta_n(z) &= \epsilon^{abc}\left[\bar d^b(z)\,\gamma^\mu C\,\bar d^a(z)^T\right]\bar u^c(z)\,\gamma_\mu\gamma_5 \\
A_\mu(y) &= \bar d(y)\,\gamma_\mu\gamma_5\, u(y),
\end{align}
where $C=i\gamma_0\gamma_2$ is the charge conjugation operator.

The correlator in momentum space is defined as
\begin{equation}
 \Pi_\mu(p,p')=\int d^4y\, d^4z\, e^{-i(q\cdot y+p\cdot z)}\,\Pi_\mu(0,y,z),\label{Eq.Pi-momentum}
\end{equation}
where the energy-momentum is conserved as shown in the diagram on the left of Fig.\,\ref{diagrams}.
The idea is to obtain the $g_A$ by relating the hadronic sector to the QCD sector through the FESR using the quark-hadron duality principle.
But first we need to isolate the contribution of $G_A$ from the other contributions.

\subsection{Hadronic sector} 

\begin{figure}
 \includegraphics[scale=0.9]{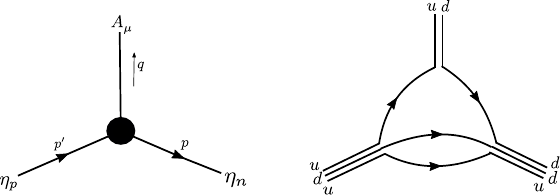}
 \caption{Feynman diagrams representing the current correlator in the hadronic sector (left) and in the QCD sector (right).}
 \label{diagrams}
\end{figure}

Inserting a complete set of intermediate states of nucleons, the correlator in momentum space for the hadronic sector leads to
\begin{equation}
 \Pi^\text{\tiny had}_\mu(p,p')=\lambda_n\lambda_p\frac{ (\slashed{p}+m_n)T_\mu(q)(\slashed{p}'+m_p)}{(p^2-m_n^2)(p'^2-m_p^2)}.
\end{equation}
This correlator is described by the left diagram in Fig.\,\ref{diagrams}, corresponding to an incoming neutron  current with momentum $p$, an outgoing axial current with momentum $q$ and an outgoing proton current with momentum $p'$.

To isolate the contribution of the axial coupling, we can decompose the correlator into the different structures relative to the Dirac matrices by tracing the correlator multiplied with the different Dirac structures, {\it i.e.}  $\mathrm{tr}[\Pi_\mu\,\Gamma]$ with  $\Gamma =I,\gamma_5,\gamma_\mu,\gamma_\mu\gamma_5,\sigma_{\mu\nu}$.
As a result, the different structures include combinations of $G_A$, $G_P$ and $G_T$. 
In particular
\begin{equation}
 \mathrm{tr}\,[\Pi_\mu(p,p')\,\gamma_\nu]=-4i\epsilon_{\mu\nu\alpha\beta}p^\alpha p'^\beta \Pi(s,s',t),
 \label{trace_corr}
\end{equation}
where $\Pi$ in the case of the hadronic correlator is
\begin{equation}
 \Pi^\text{\tiny had}(s,s',t)=\lambda_n\lambda_p\frac{G_A(t)+G_T(t)(m_n-m_p)/ m_N}{(s-m_n^2)(s'-m_p^2)},
 \label{Pi_had}
\end{equation}
with $s=p^2$, $s'=p'^2$. 
The difference in nucleon masses can be neglected in vacuum.
If different nucleon masses are assumed, it is possible to completely isolate the axial coupling part with the appropriate combination of the other correlator structures, however it is more complicated and unnecessary, even in the presence of magnetic fields.

\subsection{QCD sector}

Once the appropriate operation to isolate $g_A$ is obtained in Eq.\,(\ref{trace_corr}), it can be applied in the QCD sector.  
The projected correlator for the perturbative part of QCD in the chiral limit produces a two-loop contribution.
\begin{multline}
 \mathrm{tr}\,[\Pi^\text{\tiny pQCD}_\mu(p,p')\gamma_\nu]=  4i\epsilon_{\mu\nu\alpha\beta}N_c(N_c-1)\times\\
 \int\frac{d^4k}{(2\pi)^4}\frac{d^4k'}{(2\pi)^4}\,   \frac{32\,q^\alpha k^\beta\,k'\!\!\cdot\!(p'-k-k')}{k^2\, k'^2\,(k-q)^2 \,(p'-k-k')^2}
\end{multline}
which is described diagrammatically on the right side of Fig.\,\ref{diagrams}.
After the integration of the internal momentum in the frame $t=0$ the result is
\begin{multline}
 \Pi^\text{\tiny pQCD}(s,s',0)= 
 \frac{s^2\ln(-s/\mu^2)-s'^2\ln(-s'/\mu^2)}{(2\pi)^4\,(s'-s)}\\
 +\text{regular terms},
 \label{PI-pQCD}
\end{multline}
where $\mu$ is the $\overline{\text{MS}}$ scale.
Terms without discontinuities on the real axes or singularities are omitted because they vanish when the FESR are applied.
 
The next contribution comes from the non-perturbative sector.
Considering the operator product expansion, the next contribution in the chiral limit corresponds to dimension 3 operators: the quark condensate.
However, this term vanishes when performing the projection described in Eq.\,(\ref{trace_corr}), as well as for all diagrams with odd-dimensional operators in the chiral limit. 
Therefore, the next non-vanishing contribution comes from the dimension 4 operator, which in the chiral limit corresponds to the gluon condensate.
This diagram is complicated to handle and is not quite relevant.
Therefore, only the leading term, which corresponds to the perturbative part, will be considered.

\section{Axial coupling constant from FESR}

\begin{figure}
\includegraphics[scale=0.8]{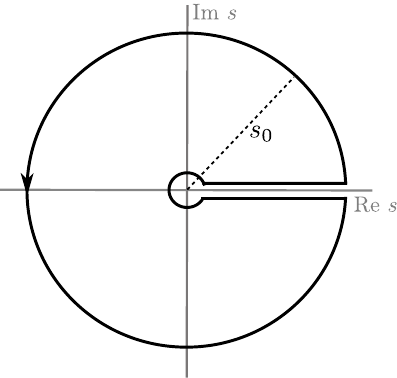}
\caption{The FESR contour.  The integration is performed over the variables $s$ and $s'$.}
\label{pacman}
\end{figure}

The FESR in this case are applied for two momentum correlators: $s$ and $s'$.
The usual procedure corresponds to integrate the correlator multiplied by an analytical kernel $K(s)$ along the  {\it pacman} contour described in Fig.\,\ref{pacman}.
The radius of the circle is the hadronic continuum threshold.
Using Cauchy's theorem, quark-hadron duality is introduced by placing the hadronic sector at the discontinuity on the positive real axis, and the QCD sector on the complex circle:
\begin{equation}
 \int_0^{s_n}\frac{ds}{\pi}\,\mathrm{Im}_s\,\Pi^\text{\tiny had}(s,s',t)=-\oint\frac{ds}{2\pi i}\Pi^\text{\tiny QCD}(s,s',t)
 \label{FESR} 
\end{equation}
where the variable $s$ is first integrated with the weight function $K(s)=1$, and where $s_n$ denotes the continuum threshold related with neutron-current. 
The subscript introduced in the imaginary part is defined as
\begin{equation}
 \mathrm{Im}_s f(s)\equiv\lim_{\epsilon\to 0}\mathrm{Im}f(s+i\epsilon).
\end{equation}
Of course, Eq.\,(\ref{FESR}) is valid in the absence of poles within the contour, otherwise the residues must be incorporated into the equation.
In particular we can see from Eq.\,(\ref{PI-pQCD}) that the pole in $s=s'$ in the denominator cancels with the numerator, so there is no singularity at all within the contour.  

Proceeding in the same way, but now with the variable $s'$, the double FESR gives
\begin{multline}
 \int_0^{s_p}\frac{ds'}{\pi} \,\mathrm{Im}_{s'}\!\!\int_0^{s_n}\frac{ds}{\pi}\,\mathrm{Im}_s\Pi^\text{\tiny had} (s,s',t)\\
 =\oint_{s_p}\frac{ds'}{2\pi i}\oint_{s_n}\frac{ds}{2\pi i}\,\Pi^\text{\tiny QCD}(s,s',t),
 \label{eq.FESR_had=QCD}
\end{multline}
where $s_p$ is the continuum threshold related with the proton-current.

Once FESR are applied to the hadronic sector in Eq. \,(\ref{Pi_had}) and to the QCD sector in Eq. \,(\ref{PI-pQCD}), after setting $t=0$, the above equation gives the relation
\begin{multline}
 g_A\lambda_n\lambda_p\,\theta(s_n-m_n^2)\theta(s_p-m_p^2)\\
=\frac{1}{48\pi^4}\left[s_n^3\,\theta(s_p-s_n)
+s_p^3\,\theta(s_n-s_p)\right].
\label{FESR_had=QCD}
\end{multline} 
In vacuum, all parameters are practically the same for protons and neutrons, so $s_p\approx s_n\equiv s_0$ and $\lambda_p\approx\lambda_n \equiv\lambda_N $
\begin{equation}
 g_A=\frac{1}{48\pi^4}\frac{s_0^3}{\lambda_N^2}.
 \label{eq.gA-vac}
\end{equation} 

The nucleon-current coupling can be obtained from the nucleon-nucleon channel.
The most appropriate channel is the nucleon-nucleon current correlator \cite{Ioffe:1983ju,Chung:1981wm,Chung:1981cc,Chung:1982rd,Chung:1984gr,Dominguez:2020sdf}
\begin{equation}
\Pi_N(x)=\langle 0|\,{\cal T}\,\eta_N(x)\,\bar\eta_N(0)\,|0\rangle.
\end{equation}
In vacuum there are only two Dirac structures, so the FESR \cite{Dominguez:2020sdf} provide two equations:
\begin{align} 
\lambda_N^2 &= \frac{s_0^3}{192\pi^4}+\frac{s_0}{32\pi^2}\langle G^2\rangle+\frac{2}{3}\langle\bar qq\rangle^2\\
 \lambda_N^2m_N &=-\frac{s_0^2}{8\pi^2}\langle\bar qq\rangle+ \frac{1}{12}\langle G^2\rangle\langle\bar qq\rangle,
 \label{Nuclear_FESR}
\end{align}
where in the last expression, vacuum dominance was considered.
The quark and gluon condensates are defined as
\begin{align}
\langle\bar qq\rangle &\equiv \frac{1}{N_f}\sum_f\langle 0| \bar q_fq_f |0\rangle\\
 \langle G^2\rangle &\equiv \langle 0 |\frac{\alpha_s}{\pi}G_{\mu\nu}^a G^{a\mu\nu}|0\rangle.
\end{align}

The axial coupling constant depends strongly on the quark and gluon condensates.
The most often used values for these operators are  $ \langle\bar qq\rangle= -(0.24\text{ GeV})^3$  and $\langle G^2\rangle= (0.33\text{ GeV})^4$.
If these values are used, the axial coupling results in $g_A=1.52$.
Recent lattice results for 2+1 flavors at the $\overline{\text{MS}}$ renormalization scale of 2 GeV obtain on average $\langle -\bar qq\rangle^{1/3}\approx 0.272\text{ GeV}$ \cite{Gubler:2018ctz}.
Close results were obtained for 2-flavor FESR, which provide $ \langle -\bar qq\rangle^{1/3}\approx 0.267\text{ GeV}$ \cite{Bordes:2010wy}.

The gluon condensate is an scale invariant quantity. 
In this case the situation is not so clear, where different estimations provide important variations: $\langle G^2\rangle^{1/4} =$ 0.3 -- 0.5 GeV \cite{Dominguez:2018zzi}

\begin{figure}
 \includegraphics[scale=0.65]{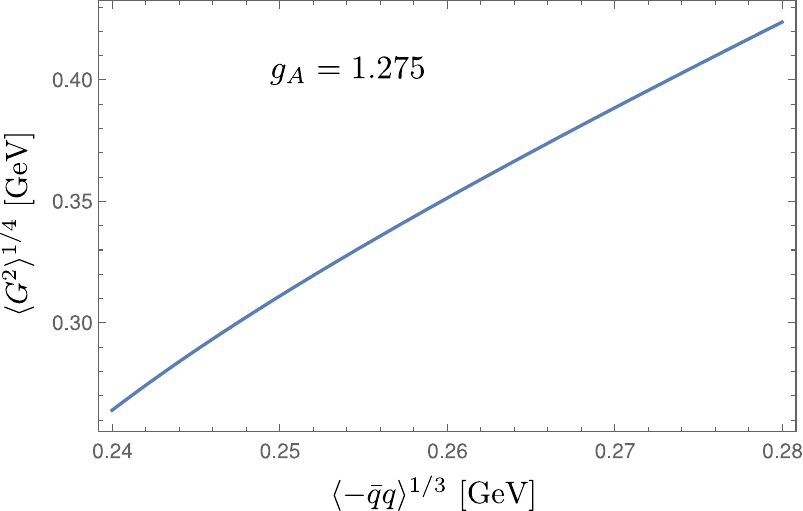}
 \caption{
 Range of values for the gluon condensate and the quark condensate for a fixed value of the axial coupling.}
 \label{gAfixed}
\end{figure}

Actual estimates of the axial coupling fluctuate around $g_A\approx 1.275$  \cite{Czarnecki:2018okw}.
Figure\,\ref{gAfixed}, shows the relation between the gluon condensate and the quark condensate by fixing the axial coupling.  
We can see that the range of values for the chiral condensate are those frequently used in the literature, and the values obtained for the gluon condensate are also in the range of accepted values.

The inclusion of more terms in the operator product expansion, as well as radiative corrections, is expected to limit solutions with more precise results.

 \bigskip

\section{Axial coupling constant at finite external magnetic field}

The FESR result $ \langle -\bar qq\rangle^{1/3}= 0.267$\,GeV will be considered, and the gluon condensate will be set to  $\langle G^2\rangle^{1/4} = 0.3775$~GeV in order to obtain $g_A= 1.275$.
This choice of condensates generates, from Eq. (\ref{Nuclear_FESR}), $\lambda_N = 0.022\text{ GeV}^3$ and $s_0=1.429\text{ GeV}^2$.

As can be seen in \cite{Ayala:2015qwa,Dominguez:2018njv,Dominguez:2020sdf}, the magnetic field can be treated by considering the expanded fermion propagator in powers of $eB$, even for light quarks.
The series is truncated by the contour integral depending on the analytical kernel used in the FESR.
The expansion of the magnetic field allows reaching values of $eB$ higher than $10\,m_\pi^2$, which is enough to obtain the phenomenology of the strong magnetic field produced in relativistic HIC experiments and in the interior of magnetars.
In this sense, the lowest order contribution is the result obtained in vacuum in Eq. (\ref{FESR_had=QCD}), but replacing the different parameters by those dependent on the magnetic field.

Aforementioned approximation is evident in the QCD sector, and the result at the lowest order is just the right-hand side of Eq.\,(\ref{FESR_had=QCD}), but what about the hadronic sector?
Let us describe the currents in the hadronic sector in another way.
The nucleon current is the nucleon field times the nucleon-current coupling. 
To reproduce the matrix elements described in Eqs. (\ref{eq.eta_p})-(\ref{eq.A}), the axial-vector current can be expressed through the nucleon fields in configuration space in the following way
\begin{align}
\eta_N(x) &= \lambda_N \psi_N(x) \label{eta_N}
\\
A_\mu(y) &= \int d^4\xi\, \bar\psi_p(\xi)T_\mu(\xi-y)\psi_n(\xi),
\end{align}
where $T_\mu(q)$ in Eq. (\ref{eq.A}) corresponds to the Fourier transform in the momentum space of the function $T_\mu(x)$ in the configuration space described in the previous equation. 
The correlator in the configuration space is therefore
\begin{equation}
\Pi_\mu(x,y,z) = -\!\!\int \!\!d^4\xi~ e^{i\Phi(x,\xi)} \, S_p^B(x-\xi)\, T_\mu(\xi-y)\, S_n^B(\xi-z)
\end{equation}
where the magnetic field dependent proton propagator is described by the local part multiplied by the Schwinger phase.
The neutron propagator in the presence of the magnetic field contains the anomalous magnetic moment contribution.
The definition of the correlator in momentum space in Eq. (\ref{Eq.Pi-momentum}) is not arbitrary. 
In fact, choosing the frame $x=0$, the Schwinger phase disappears if the Fock-Schwinger gauge ${\cal A}_\mu(x) = -\frac{1}{2}F_{\mu\nu}x^\nu$ is considered for the external electromagnetic vector field.
The correlator in momentum space is therefore
\begin{align}
\Pi_\mu(p',p)
&=-S_p^B(p')T_\mu(q)S_n^B(p).
\end{align}
Considering the expanded propagators in magnetic field power series \cite{Dominguez:2020sdf}, it is not difficult to see that, in the sum rule considered in Eq. (\ref{FESR}), only the lowest order term in the expansion will survive if we keep the same $T_\mu$ structure described in Eq. (\ref{T_mu}).
However, when a magnetized medium is present, the overall structure of $T_\mu$ must incorporate the external electromagnetic tensor $F_{\mu}$ which contributes to other structures, dividing the axial contribution of the form factor as
\begin{equation}
 G_A\gamma_\mu\to G_A^\parallel\gamma_\mu^\parallel
    + G_A^\perp\gamma_\mu^\perp
    + \tilde G_A F_{\mu\nu}\gamma^\nu,
\end{equation}
as well as the other form factor terms in Eq. (\ref{T_mu}) will be divided into several substructures.
The difference between $G_A^\parallel$ and $G_A^\perp$ will be of order $(eB)^2/s_0$.
Since we are considering the lowest term of the expansion, there will be no difference between $G_A^\parallel$ and $G_A^\perp$.

\begin{figure}
\includegraphics[scale=0.6]{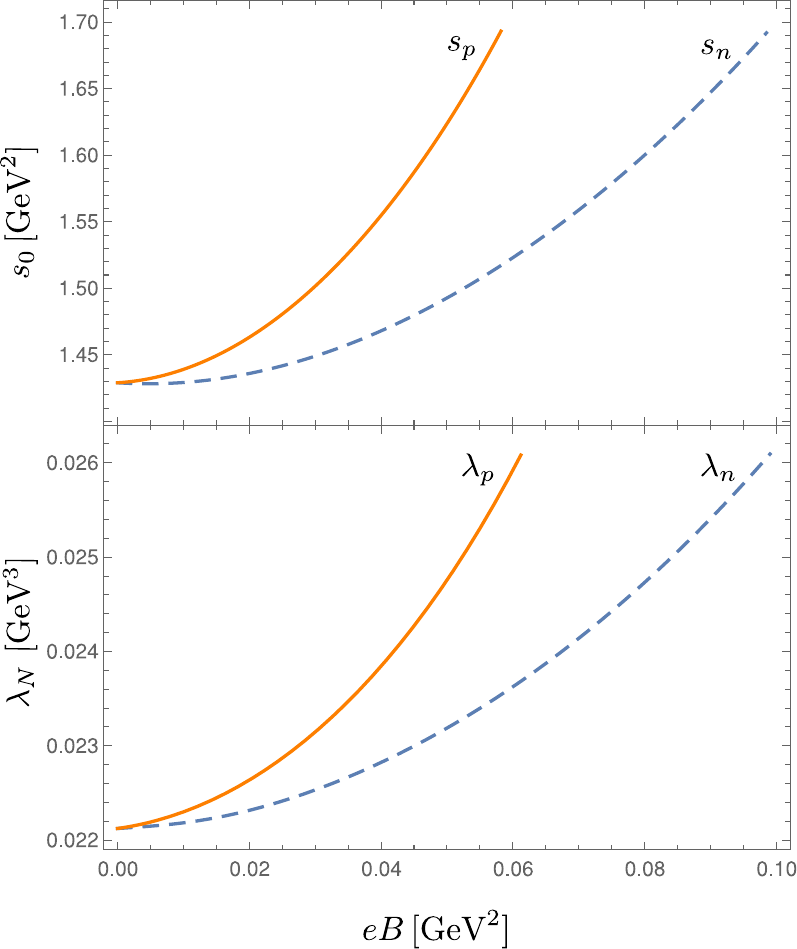}
\caption{Hadronic threshold (upper panel) and nucleon current coupling (lower panel) for proton (solid line) and neutron (dashed line)}
 \label{fig.s0-lambda}
\end{figure}

With all the above considerations, the axial coupling constant at finite magnetic field is given by Eq.\,(\ref{FESR_had=QCD}).
The hadronic thresholds and current-nucleus couplings are obtained from \cite{Dominguez:2020sdf}, changing the values of the quark and gluon condensates in vacuum to the values defined at the beginning of this section.
The resulting hadronic thresholds and nucleon-current couplings are plotted in Fig.\,\ref{fig.s0-lambda}.
The first thing to note is that $s_p>s_n$, and therefore the axial coupling constant of the relation in Eq.\,(\ref{FESR_had=QCD}) can be written as
\begin{align}
g_A=\frac{1}{48\pi^4}\frac{s_n^3}{\lambda_p\lambda_n}.
\end{align}

\begin{figure}
\includegraphics[scale=0.65]{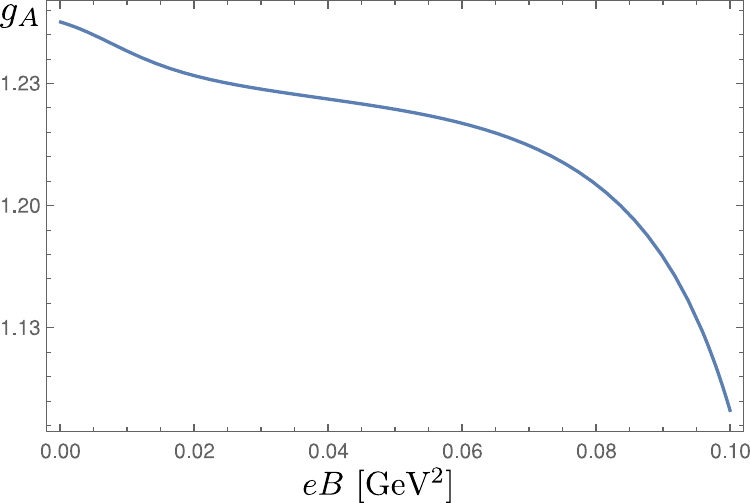}
\caption{Axial coupling constant as a function of an external magnetic field.
}
 \label{fig.gA_B}

\end{figure}

Both the hadronic threshold of the nucleons and the nucleon-current couplings increase with the magnetic field as can be seen in Fig.\,\ref{fig.s0-lambda}.
The evolution of the axial coupling constant as a function of external magnetic field is shown in Fig.\,\ref{fig.gA_B}. 
The axial coupling constant decreases, as the magnetic field increases. 
For lowest values of the magnetic field, this decrease is linear.

The decrease in axial nucleon coupling is due to flavor asymmetry.
Since there is competition between the smaller hadronic threshold of nucleons and the nucleon-current couplings, then the proton coupling dominates.

\section{Conclusions and discussion}

The effects of an external uniform magnetic field on the axial coupling constant were obtained by finite energy sum rules.
By isolating the axial structure in the proton-axial-neutron current correlator, and performing a double contour integral, it is possible to match the hadronic part with the perturbative QCD part.
The relevant parameters in this case, the nucleon-current couplings and the hadronic thresholds are obtained through the nucleon-nucleon correlator at finite external magnetic field calculated in \cite{Dominguez:2020sdf}.

The axial coupling is then proportional to the neutron threshold cubed and inversely proportional to the proton-current and neutron-current couplings.
Both thresholds and couplings are increasing quantities as a function of magnetic field, but the proton-current coupling dominates, and the axial coupling constant decreases with $B$.
The change with the magnetic field about 10\% for $eB=0.1\text{ GeV}^2$.
In particular, the factor $1+3g_A^2$, which is proportional to the neutron decay width as well as the neutrino emissivity in the Urca process, decreases by 16\%.
This is an effect to take into consideration. 
Unfortunately, since this is the first attempt to find the magnetic evolution of the axial coupling constant, there are no other solutions to compare with.
The axial coupling of pions with constituent quarks has been calculated for $eB\sim 0.01$\,GeV$^2$, presenting a linear behavior \cite{Braghin:2018drl}.
This could be a related form factor but not the same, so it is necessary to verify nucleon axial coupling constant under an external magnetic field using other models and techniques.

It is interesting to see what happens in high temperature scenarios, such as the relativistic HIC experiments and high density scenarios such as in magnetars.
Apparently temperature and baryon density effects tend to reduce axial coupling, but it is not clear what may happen with temperature or density effects in combination with the external magnetic field.
The case of baryon density and magnetic field effects will be addressed soon.

\bigskip\bigskip

\begin{acknowledgments}
I would like to thank Ces\'areo Dom\'inguez and Marcelo Loewe for their fruitful discussions.
This article was funded by Fondecyt under grants 1190192, 1200483 and 1220035.
\end{acknowledgments}


%

\end{document}